%% file: main.tex
\documentclass[%
 reprint,
 amsmath,amssymb,
 aps,
pra,
]{revtex4-2}

\usepackage{dcolumn}
\usepackage{graphics, graphicx}
\usepackage{subfig}
\usepackage{multirow}
\usepackage{caption}
\usepackage{physics}
\usepackage{mathrsfs}
\usepackage{color}
\usepackage{xcolor}

\usepackage{float}
\makeatletter
\let\newfloat\newfloat@ltx
\makeatother

\usepackage{algorithm}
\usepackage{algorithmic}
\usepackage{bbm}

\begin{document}

\preprint{APS/123-QED}

\title{Noisy Tensor Ring approximation for computing gradients of Variational Quantum Eigensolver for Combinatorial Optimization }

\author{Dheeraj Pedireddy, Utkarsh Priyam, and Vaneet Aggarwal}
\affiliation{%
Purdue University, West Lafayette IN 47906
}%
\email{\{dpeddire, upriyam,  vaneet\}@purdue.edu}


\begin{abstract}
Variational Quantum algorithms, especially Quantum Approximate Optimization and Variational Quantum Eigensolver (VQE) have established their potential to provide computational advantage in the realm of combinatorial optimization. However, these algorithms suffer from classically intractable gradients limiting the scalability. This work addresses the scalability challenge for VQE by proposing a classical gradient computation method which utilizes the parameter shift rule but computes the expected values from the circuits using a tensor ring approximation. The parametrized gates from the circuit transform the tensor ring by contracting the matrix along the free edges of the tensor ring. While the single qubit gates do not alter the ring structure, the state transformations from the two qubit rotations are evaluated by truncating the singular values thereby preserving the structure of the tensor ring and reducing the computational complexity. This variation of the Matrix product state approximation grows linearly in number of qubits and the number of two qubit gates as opposed to the exponential growth in the classical simulations, allowing for a faster evaluation of the gradients on classical simulators. 
\end{abstract}

\maketitle

\input{s1_intro}

\input{s2_problem}

\input{s3_method}
\input{s4_experiments}
\input{s5_conclusion}


\appendix
\section{Commonly used gates}
The matrix representation of some of the commonly used gates in the manuscript are listed below:
$$
R_x(\theta) = 
\begin{bmatrix}
cos(\theta/2) &  -isin(\theta/2)\\
-isin(\theta/2) & cos(\theta/2)
\end{bmatrix},$$  
$$R_y(\theta) = 
\begin{bmatrix}
cos(\theta/2) &  -sin(\theta/2)\\
sin(\theta/2) & cos(\theta/2)
\end{bmatrix},$$  $$  R_z(\theta) = 
\begin{bmatrix}
e^{-i\theta/2} &  0\\
0 & e^{i\theta/2}
\end{bmatrix}
$$

$$
H =\frac{1}{\sqrt{2}}
\begin{bmatrix}
 1&  1\\
1 & -1
\end{bmatrix}
$$

$$
CNOT = 
\begin{bmatrix}
1 & 0 & 0 & 0\\
0 & 1 & 0 & 0\\
0 & 0 & 0 & 1\\
0 & 0 & 1 & 0
\end{bmatrix}
$$

$$
R(\alpha, \beta, \gamma) = 
\begin{bmatrix}
cos(\alpha/2) & -e^{i\gamma}sin(\alpha/2) \\
e^{i\beta}sin(\alpha/2) & e^{i\beta + i\gamma}cos(\alpha/2)
\end{bmatrix}
$$

\bibliography{references}

\end{document}

%% file: s1_intro.tex
\section{Introduction} \label{sec:1}
Quantum computing has been far touted for its potential to solve some complex problems much more efficiently than the classical computers \cite{10.1145/237814.237866, shor1999polynomial}. Although the fruition of the idea is further into the future, researchers have been exploring the real-time applicability of the current generation quantum computers. Most of the quantum processors in their current state are severely limited in the number of qubits, noise levels and inefficient error mitigation techniques, calling for a class of algorithms robust to the noise and error. Variational Quantum Algorithms (VQA) have been studied widely for their resilience to the noise from decoherence making them an ideal choice of algorithms for various applications on gate-based Noisy Intermediate Scale Quantum (NISQ) devices. Two such algorithms of prominence, Variational Quantum Eigensolver (VQE) and Quantum Approximate Optimization Algorithm (QAOA) evaluate the expected energy of a state resulting from a short parameterized circuit (frequently referred to as ansatz) with respect to an observable defined by a given problem. A classical outer-loop optimizer tries to find the optimal circuit parameters that minimize the expected energy. While QAOA implements a fixed ansatz inspired from adiabatic quantum computing, VQE utilizes a variable ansatz offering flexibility to engineer the ansatz based on the hardware constraints and the problem at hand. This work chooses to focus on VQE, inspired by the recent advances of variable ansatz in quantum machine learning \cite{cerezo2021variational}. VQE, initially developed by Peruzzo et al. \cite{peruzzo2014variational}, has seen a number of applications in condensed matter physics \cite{van2020rechargeable, bravo2020scaling, continentino2021key}, quantum chemistry \cite{deglmann2015application, williams2018free, heifetz2020quantum} and quantum mechanics \cite{miceli2019effective, banuls2020simulating}.

Optimization is one of the frontrunners among the applications being studied for potential quantum advantage from VQE and adjacent algorithms \cite{amaro2022filtering, moll2018quantum, nannicini2019performance}. Combinatorial optimization is a class of problems of practical relevance with applications spanning across transportation, logistics, manufacturing etc. Studies have indicated that the exponentially growing state space and quantum entanglement can improve the chances of finding the right solution with a potential speedup \cite{abrams1999quantum, harrow2009quantum}. Even minor improvements  to optimization problems from quantum algorithms can potentially have a large impact on the society. In the context of VQE, a multi-qubit Hamiltonian is prepared with its ground state encoding the solution of the optimization problem and the algorithm optimizes their parameters to minimize the energy of the Hamiltonian. The algorithm has been extended to use filtering operators \cite{amaro2022filtering} and iterative approaches \cite{liu2022layer}, to improve the performance with combinatorial optimization. The approach has also been validated on several practical applications using optimization (e.g., Job Shop Scheduling \cite{amaro2022case}, Vehicle Routing \cite{atchade2020using}) 

\begin{figure*}[!t]
\centering
	\subfloat{\includegraphics[keepaspectratio,width=0.5 \textwidth]{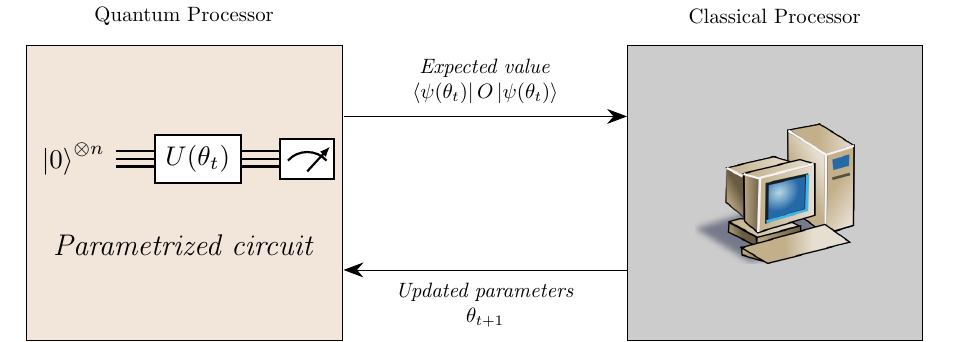}\label{fig:vqc}}
	\subfloat{\includegraphics[keepaspectratio,width=0.5 \textwidth]{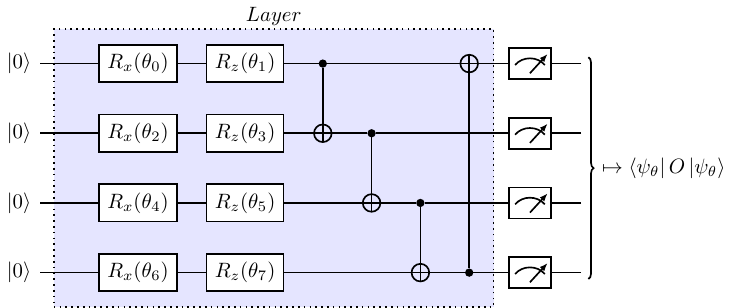}\label{fig:vqe}}

\caption{(Left) High-level architecture of a quantum-classical hybrid algorithm with the quantum processor implementing a circuit parameterized by $\theta_t$ at iteration $t$, which is used to compute the expected value in regards to the observable $O$. The classical processor processes the gradients of the expected value with respect to the parameters $\theta$ and update them by gradient-descent like methods  (Right) An example of parameterized circuit with 4 qubits with single qubit gates parameterized by $\theta_i$ and the entanglement is encoded into the quantum state using the $CX$ gates. A layer comprises of a certain set of gates and each layer can be repeated $D$ times based on the complexity of the problem }%
\label{fig:vqe-sum}%
\end{figure*}

Despite promising prospects, VQAs and more broadly quantum circuits are hindered by a plethora of problems in the current era of quantum computing, with the primary forces of impedance being the limited number of qubits, physical cost of implementation of quantum circuits and decoherence noise. Hybrid algorithms also suffer from the asymmetric scaling of quantum and classical resources with the circuit execution scaling linearly in number of qubits and circuit depth and the classical gradient evaluation scaling exponentially. Note that the gradients of the variational parameters in VQAs were evaluated using either automatic or numeric differentiation until Schuld et al. \cite{schuld2019evaluating} formalized the notion for gradients computed on quantum hardware popularized as the parameter shift rule. This method estimates the gradients by computing the energy of the wave functions generated by identical circuits with the parameter for which the gradient is to be estimated, shifted by certain values. Parameter shift rule alleviates the imbalance in the scalability, albeit at the cost of executing a much larger number of quantum circuits than the other methods. Given the inconsistency in evaluating the expected values from circuits due to decoherence and inefficient error mitigation techniques on top of the statistical noise from measurement, a larger number of circuits can lead to inaccurate results.

In order to address the issues of scalability, accuracy and cost of execution, this manuscript proposes a classically simulated quantum circuit execution method that approximates the initial and intermediate quantum states using a low-rank tensor ring (TR) to compute the expected energy, which in turn are used in approximating the gradients of a VQE. Built upon the Matrix Product State (MPS) approximation of many body quantum states \cite{vidal2003efficient}, the tensor ring VQE (TR-VQE) formulates a combinatorial optimization in the same way as a naive VQE, using parameter shift rule to compute the gradients. However, the expected values of the shifted circuits used to compute the gradients are evaluated by approximating the initial quantum state with a TR as opposed to MPS, where the single qubit and two qubit gates corresponding to the circuit ansatz are evaluted using tensor contractions. It must be noted that while a single qubit gate does not change the structure of the tensor network, a two qubit gate contracted with the two corresponding tensors can alter the network by increasing the tensor size or its rank. The proposed method retains the tensor ring structure and rank by truncated singular value decomposition of the higher order tensor resulting from the application of two-qubit gate. The consistent low-rank structure allows for an exponential speedup with respect to the number of qubits and circuit depth, compared to the MPS approximation and the brute force approximation with full state vector. This truncation however, induces a noise in the circuit executions similar to the decoherence in actual quantum computers. Therefore, classically simulating a noisy quantum computer instead of a perfect quantum computer only scales linearly in the number of qubits and circuit depth \cite{zhou2020limits}. MPS representation tries to simulate ideal quantum computation without noise but literature suggests that the noise in the current generation quantum computers limits the amount of entanglement that can be built into a quantum state. Given the computational cost of simulating ideal quantum computers, this may not be an ideal prospect since they are not representative of the noisy quantum computations. Moreover, given the robustness of VQAs to noise, this kind of noisy simulation with the benefits of scalability can be specifically useful for machine learning and optimization. Furthermore, Liu et al. \cite{liu2022noise} highlights that the presence of noise in VQAs can naturally help the optimizer avoid saddle points. We posit that this advantage extends to the TR-VQE as well due to the induced noise. The proposed method is validated on multiple instances of max-cut problem compared against F-VQE \cite{amaro2022filtering} and a naive VQE using parameter shift rule. The expected values of the circuit for the benchmarks are computed using simulations implementing a non-noisy MPS approximation highlighting the improved performance of noisy TR approximation over MPS approximation. 

\begin{figure}[!t]
 \centering
 \includegraphics[height=5cm, width=8cm]{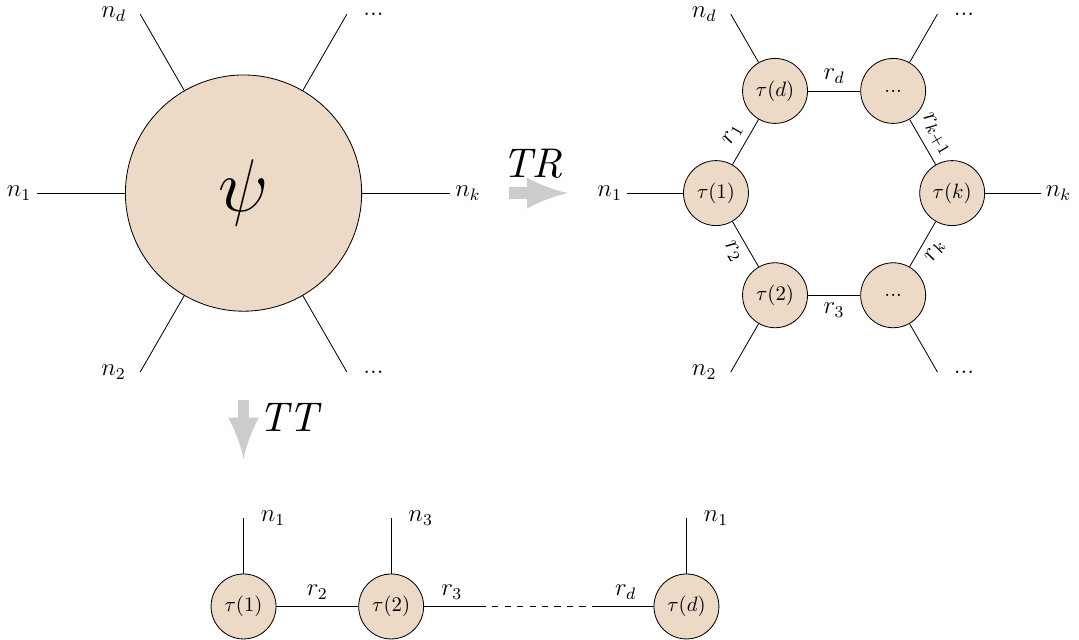}
 \caption{\textbf{Illustration of a tensor train and tensor ring decomposition} A higher order tensor with order-$d$ and dimensionality $n_k$ at the $k$-th edge is decomposed into (i) a Tensor Ring with all the $d$ tensors multiplied at the indices denoted by $r_k$ and (ii) a Tensor Train whose border tensors are constrained to have an order of 2 while the internal tensors have an order of 3}
 \label{fig:trd}
\end{figure}

The rest of the manuscript is organized as follows: Section \ref{sec:1-1} recounts the existing literature related to the use of Tensor networks in approximating quantum circuits and applications in QML. Section \ref{sec:2-2} formulates the notion of VQE to solve maximum cut problem introduced in Section \ref{sec:2-1}. Section \ref{sec:3-1} discusses the proposed method used to compute the gradients of a variational quantum circuit using the TR approximation of a quantum state and Section \ref{sec:3-2} addresses the complexity analysis of the proposed method. The numerical simulations are explained in Section \ref{sec:4} followed by a discussion on limitations and future direction in Section \ref{sec:5} 

\subsection{Related Work}\label{sec:1-1}
Since its inception, the tensor network approach has been much more widely explored in the context of classical simulation of quantum computations, compared to the brute-force statevector simulation 
or other graphical and distributed methods \cite{de2007massively, boixo2017simulation}. Matrix Product states especially were widely regarded for their ability to efficiently represent moderately entangled quantum many body states \cite{vidal2003efficient}. The idea has been further extended to techniques that efficiently simulate quantum circuits \cite{markov2008simulating} by contracting tensor networks at a fraction of cost of the statevector simulation which holds the full $2^N$ sized vector. Building upon the literature several variations have emerged for specific cases like Projected Entangled Pair States (PEPS) for two-dimensional circuits \cite{verstraete2008matrix} and Tree Tensor networks (TTN) for circuits with tree-like connectivity \cite{shi2006classical} and Multi-scale Entanglement Renormalization Ansatz (MERA) \cite{vidal2007entanglement} etc.

Note that the naive MPS-based circuit simulation (which will be referred to as non-noisy MPS approximation in this manuscript) as formulated in \cite{markov2008simulating} and widely implemented across quantum computing platforms like Qiskit, do not efficiently encode circular entanglement from first to last qubits. Further, any application of two-qubit gate contractions result in increasing tensor size which in turn increases the computational complexity as the number of two-qubit gates in the circuit increases. To circumvent this shortcoming, Zhou et al. \cite{zhou2020limits} proposed a truncated MPS approximation to simulate noisy quantum computers, which demonstrates a linear complexity in number of qubits and circuit depth.
The noisy simulation addresses the issue of increasing tensor size by approximating the larger tensor after the application of a two qubit gate with tensors of smaller size. The higher order tensor is decomposed into two lower order tensors by truncated singular value decomposition. This approximation preserves the tensor sizes after the application of each gate unlike in the previous iterations of MPS-based simulation.

A number of quantum-inspired tensor network methods have been explored in the machine learning literature for supervised learning. Huggins et al. \cite{huggins2019towards} implements MPS and Tree Tensor Network models to solve binary classification. Other tensor network based methods using PEPS and MPS were demonstrated to be effective in image classification tasks \cite{cheng2021supervised, efthymiou2019tensornetwork, stoudenmire2016supervised}. The aforementioned literature mostly explores quantum-inspired classical machine learning techniques but very few works have probed into the utility of tensor networks in augmenting quantum machine learning techniques. Peddireddy et al. \cite{peddireddy2022tensor} extends the singular value thresholding method from Zhou et al. \cite{zhou2020limits} to tensor rings implemented with variational quantum classifiers demonstrating the scalability and improved performance over non-noisy MPS approximation. Tensor rings also encode circular entanglement more efficiently than MPS due to the ring structure. While Zhou et al. \cite{zhou2020limits} evaluates the approximated expectations using noisy MPS representation, they do not explore the notion of extending it to computing gradients of variational circuits. Therefore, the application of noisy circuit simulation to scale the classical optimization loop of VQE is still an open problem. Furthermore, extending this approximation method from MPS to tensor rings can also improve representability. This work builds up on \cite{peddireddy2022tensor} and \cite{zhou2020limits} by adapting the noisy tensor ring representation to compute the approximate gradients of the parameters of a variational quantum eigensolver using the parameter-shift rule. Although the proposed TR based representation computes less accurate gradients than non-noisy MPS based representations owing to the additional information that is removed in the form of truncated singular values, TR based approach scales much more efficiently.

%% file: s2_problem.tex
\section{Problem Setup}\label{sec:2}

\subsection{Max-Cut Optimization Problem}\label{sec:2-1}

This section will briefly introduce the maximum cut (max-cut) problem and its mathematical notion in the context of quantum computers. Max-Cut is an NP-hard binary optimization problem with a history of applications in statistical physics, VLSI design and clustering etc. Given an undirected graph $G = (V,E)$, with $V \text{and } E$ representing the nodes and edges of the graph, the problem aims to maximize the summed weights of the edges that are cut by grouping the nodes of the graph into two subsets by choosing the optimal subgroups. 

The mathematical definition follows the QUBO formulation \cite{glover2018tutorial}: a graph of $n$ nodes with the weights of the edges given by $w_{ij}$ for $(i,j) \in E$. The nodes of the graph are cut into two subgroups labelled $+1$ and $-1$. The problem attempts to maximize the objective function $C(x)$ given by the sum of the weights of the edges connecting the nodes in $+1$ to the nodes in $-1$ which assumes the form: 

\begin{equation}\label{eq:one}
    C(x) = \sum_{i,j} w_{ij} x_i (1 - x_j) 
\end{equation}

where $x \in \{0, 1\}^{n}$ and $(i,j) \in E$. The bitstring $x$ corresponds to an instance of the grouping schema where $x_i = 0 \text{ or }1$ represents the i-th node being assigned to the subgroup +1 or -1 respectively. In order to find the solution to the given objective function with a quantum computer, we construct an Ising Hamiltonian \cite{lucas2014ising} corresponding to the function by substituting $x_i$ with its matrix transformation $\frac{I- Z_i}{2}$ where $Z_i$ are the Pauli $Z$ operators that act on qubit $i$ and $I$ is the identity matrix:

\begin{equation}\label{eq:two}
    C(x) = \sum_{i,j} \frac{1}{4} w_{i,j} (I - Z_i) (I + Z_j)
\end{equation}

\begin{equation}\label{eq:three}
    C(x) = \frac{1}{2}\sum_{i<j} w_{ij} - \frac{1}{2}\sum_{i<j}Z_i Z_j
\end{equation}

Essentially, maximizing the objective of the given optimization problem is equivalent to minimizing the energy of Ising Hamiltonian given by:
\begin{equation}\label{eq:four}
    \mathcal{H} = \sum_{i,j} w_{i,j} Z_i Z_j
\end{equation}

whose ground state corresponds to the solution of the optimization.The full Hamiltonian $\mathcal{H} \in \mathbb{C}^{2^n}$ is never constructed explicitly but is represented using a combination of the Pauli Z operators.

\begin{algorithm}[t]
\caption{Solving Max-Cut Optimization using VQE}
\begin{flushleft}
\textbf{Input}: Number of nodes $K$, Edge weights $w_{ij}$, Variational Circuit $U(\theta)$, Initialization $\theta_{0}$, number of iterations ($T$), Learning rate ($\alpha$) \\
\textbf{Output}: Updated weights ($\theta_{T}$)
\end{flushleft}

\begin{algorithmic}[1]

\STATE Define the Ising Hamiltonian $\mathscr{H}$ from the problem graph as specified in Equation (\ref{eq:four})
$$
\mathcal{H} = \sum_{i,j} w_{i,j} Z_i Z_j
$$
\FOR{$t = 1,2 \cdots T$}
\STATE Prepare the initial state of $N = K-1$ qubits in  $\ket{0}^{\otimes N}$
\STATE Apply the unitary transformation using quantum gates defined by the circuit $U(\theta_{t-1})$ to the initial state
$$
\ket{\psi(\theta_{t-1})} = U(\theta_{t-1})\ket{0}^{\otimes N}
$$
\STATE Compute the energy expectation of the state with respect to the observable Hamiltonian on a quantum device
$$
\expval{H(\theta_{t-1})} = \bra{\psi(\theta_{t-1})}\mathcal{H}\ket{\psi(\theta_{t-1})}
$$
\STATE Compute the gradients of weights with respect to the energy expectation using parameter shift rule for all $i$ given in Equation \ref{eq:seven}
$$
\pdv{\expval{H(\theta_{t-1})}}{\theta^i}_{\theta_{t-1}} \quad = \frac{\expval{H(\theta_{t-1} + \frac{\pi}{2} \mathbbm{1}_i)} - \expval{H(\theta_{t-1} - \frac{\pi}{2} \mathbbm{1}_i)}}{2}
$$
\STATE Update weights:
$$
\theta_t = \theta_{t-1} - \alpha \pdv{\expval{H(\theta_{t-1})}}{\theta^i}_{\theta_{t-1}}
$$
\ENDFOR
\end{algorithmic}
\label{algo:one}
\end{algorithm}

\subsection{Variational Quantum Eigensolver}\label{sec:2-2}

VQE is one of the algorithms that utilizes parameterized quantum circuits to solve for an approximate solution of combinatorial optimization problems. Unlike QAOA, VQE does not enforce any constraints on the circuit ansatz and therefore can be altered to suit the hardware that it's being implemented on. The optimization problem is first translated to a qubit Hamiltonian $\mathcal{H}$ whose eigenvalues correspond to the costs of various solutions with the ground state being associated with the optimal solution of the problem. A quantum circuit with parameterized unitary rotations denoted by $U(\theta)$ is applied to an initial state $\ket{\psi_0}$ (generally chosen to be the basis state $\ket{0}^{\otimes n}$) resulting in a trial wavefunction.
\begin{equation}\label{eq:five}
    \ket{\psi(\theta)} = U(\theta)\ket{\psi_0}
\end{equation}

Here, $U(\theta)$ represents a chosen ansatz $U$ with variational parameters given by $\theta$. The energy landscape of the Hamiltonian can be traversed using this wavefunction to estimate the expected energy. We choose the notation $\expval{H(\theta)}$ to represent the expectation value of $\ket{\psi(\theta)}$ with respect to the observable Hamiltonian $\mathcal{H}$.
\begin{equation}\label{eq:six}
    \expval{H(\theta)} = \bra{\psi(\theta)}\mathcal{H}\ket{\psi(\theta)}
\end{equation}

The algorithm then updates the variational parameters of the circuit employing an outer loop optimizer using gradient descent or other adjacent methods. The process is repeated until we arrive at a sufficiently low energy. The quality of the solution at the $t$-th iteration is evaluated using the approximation ratio which is defined as follows:
\begin{equation}\label{eq:approx}
    \alpha = \frac{M-\expval{H(\theta_t)}}{M-m}
\end{equation}
where $M$ represents the maximum possible Hamiltonian value and $m$ the minimum. In other words, $\alpha=1$ represents the optimal solution, and $\alpha=0$ represents making no cuts.

Most of the variational quantum algorithms including VQE are implemented as hybrid models that compute the expected value of the observable on a quantum computer while calculating gradients and updating the weights on a classical computer. The fundamental mechanics of the VQE algorithm is illustrated in Figure \ref{fig:vqe-sum}. Following the parameter shift rule \cite{schuld2019evaluating, mitarai2018quantum}, when the variational parameters are components of a single qubit rotation gate, the gradient takes the following form:

\begin{equation}\label{eq:seven}
    \pdv{\expval{H(\theta)}}{\theta^i} = \frac{1}{2}[\expval{H(\theta + \tfrac{\pi}{2} \mathbbm{1}_i)} - \expval{H(\theta - \tfrac{\pi}{2} \mathbbm{1}_i)}]
\end{equation}

Given the choice of ansatz, we choose a circuit  that only comprises $CX$ ($CNOT$) gates and single qubit rotation gates which form a universal gate set, thus simplifying the gradients to the closed form given in Equation \ref{eq:seven} where  $\theta^i$ is the $i$-th element of $\theta$,  $\expval{H(\theta)}$ corresponds to the energy of the Hamiltonian $\mathcal{H}$ with respect to the wavefunction generated by the circuit $U(\theta)$ and $\mathbbm{1}_i$ is a one-hot vector with the $i$-th value as 1.

\begin{algorithm}[t]
\caption{Computing energy expectation using TR representation}
\begin{flushleft}
\textbf{Input}: Circuit ansatz as an ordered set of one and two qubit gates ($\mathbbm{U}$), tensor ring bond ($\chi$), Observable as a combination of Pauli Operators, Number of qubits ($N$)\\
\textbf{Output}: Energy expectation of the observable matrix
\end{flushleft}
\begin{algorithmic}[1]
\STATE Compute $N$ tensors $\mathbbm{1}_{(1,1,1)}$ of dimension $\chi \times \chi \times 2$ corresponding to the initial state $\ket{0}^{\otimes N}$
\FOR{each gate $U$ in $\mathbbm{U}$}
\IF{$U \in \mathbbm{C}^{2\times2}$}
\STATE Transform the TR following the procedure from the equation (\ref{eq:nine})
\ELSIF{$U \in \mathbbm{C}^{4\times4}$}
\STATE Transform the TR following the procedure from equations (\ref{eq:ten})-(\ref{eq:fourteen})
\ENDIF
\ENDFOR
\STATE Compute the expected overall energy as a weighted sum of components evaluated as shown in equation (\ref{eq:seventeen})
\end{algorithmic}
\label{algo:two}
\end{algorithm}

%% file: s3_method.tex
\section{Methodology}\label{sec:3}
\subsection{Computing gradients using Tensor Rings}\label{sec:3-1}
Since the gradients of VQE can be computed by implementing quantum circuits, it is crucial to be able to carry out the circuits efficiently. Although the parameter-shift method is faster than the automatic differentiation, it requires a quantum processor to run  three identical ansatz with different parameters numerous times to arrive at the gradients (More discussion on this is provided in section \ref{sec:3-2}). This could present an impediment given the limited availability of quantum computers and the cost of each implementation. Therefore, it is essential to study the utility of classical simulation of quantum circuits in assisting the optimization procedure.

Tensor networks have been shown to be effective in approximating quantum many body systems and are thus a strong contender among the methods for efficiently simulating quantum circuits. A tensor network can be easily understood via  Penrose diagrams or Tensor Network diagrams where each diagram corresponds to a graph of multiple nodes with each node representing a tensor. A tensor is a multidimensional array of with its order denoting the number of its dimensions or edges. A popular approximation strategy for quantum systems involve Matrix Product States(MPS) or Tensor Trains (TT), a class of tensor networks that aim to represent a higher order tensor as a chain of order-3 tensors (See Figure \ref{fig:trd}). This representation has the advantage of the topological similarity with a multi-qubit system where each tensor corresponds to a single qubit and the contraction between the tensors encodes the entanglement between the qubits. However, TTs are limited in their flexibility and representation ability due to the constraint on their border rank. Since the border ranks are much lower than the inner ranks, this representation may not be optimal for some specific quantum systems. Also, an optimal TT representation greatly depends on the order of the products restricting the choice of ansatz. Note that the border rank constraints present the same hindrances in the application of TTs to classical datasets as well. In order to ameliorate these issues, researchers in the area of classical machine learning have adopted Tensor Rings(TR) to represent the data \cite{wang2017efficient, wang2018wide}. TR structures relaxes the rank constraints on the border tensors increasing the expressibility of the tensors. TR decomposition multiplies the tensors circularly therefore removing the variance to permutations of the multiplicative order. Notable advantages of TR representation with respect to quantum states involves flexibility in the choice of the ansatz. To explain this further, let us assume a circuit similar to the one shown in Figure \ref{fig:vqe} where entanglement was introduced between the first and the last qubits using a $CX$ between the said qubits. TR representations are a better fit to encode this kind of cyclic entanglement, therefore improving the choice set of ansatz for the problem.


\begin{algorithm}[t]
\caption{Solving Max-Cut Optimization using TR-VQE}
\begin{flushleft}
\textbf{Input}: Number of nodes $K$, Edge weights $w_{ij}$, Variational Circuit $U(\theta)$, Initialization $\theta_{0}$, number of iterations ($T$), Learning rate ($\alpha$) \\
\textbf{Output}: Updated weights ($\theta_{T}$)
\end{flushleft}

\begin{algorithmic}[1]

\STATE Prepare the Ising Hamiltonian $\mathcal{H}$ as specified in Equation (\ref{eq:four})
$$
\mathcal{H} = \sum_{i,j} w_{i,j} Z_i Z_j
$$
\FOR{$t = 1,2 \cdots T$}
\STATE Compute the energy expectation $\expval{H(\theta_{t-1} \pm \frac{\pi}{2} \mathbbm{1}_i) )}$ using Algorithm \ref{algo:two} with respect to the unitary circuits $U(\theta_{t-1} \pm \frac{\pi}{2} \mathbbm{1}_i) )$ and the Hamiltonian $\mathcal{H}$

\STATE Compute the gradients of weights with respect to the energy expectation using parameter shift rule for all $i$ given in Equation (\ref{eq:seven})
$$
\pdv{\expval{H(\theta_{t-1})}}{\theta^i}_{\theta_{t-1}} \quad = \frac{\expval{H(\theta_{t-1} + \frac{\pi}{2} \mathbbm{1}_i)} - \expval{H(\theta_{t-1} - \frac{\pi}{2} \mathbbm{1}_i)}}{2}
$$
\STATE Update weights:
$$
\theta_t = \theta_{t-1} - \alpha \pdv{\expval{H(\theta_{t-1})}}{\theta^i}_{\theta_{t-1}}
$$
\ENDFOR
\end{algorithmic}
\label{algo:three}
\end{algorithm}

A quantum state $\ket{\psi} \in \mathbb{C}^{2^N}$ can be approximated by a tensor ring with N tensors (corresponding to N qubits) circularly multiplied with each tensor denoted by $\tau(n)$.

\begin{equation}\label{eq:eight}
    \ket{\psi} =  \sum_{i_1 \ldots i_N}\sum_{r_1 \ldots r_N} \tau(1)_{r_N r_1}^{i_1} \tau(2)_{r_1 r_2}^{i_2} \ldots \tau(N)_{r_N r_1}^{i_N} \ket{i_1 i_2 \ldots i_N} 
\end{equation}

Here, free indices $i_n \in \{0, 1\}$ span the $2^N$ dimensional Hilbert space corresponding to the quantum state whereas $r_n$ represent the bond indices (indices connecting the tensors) with rank $\chi_n$, which determines the quality of the approximation with entangled states i.e., higher values of $\chi_n$ are better able to represent strongly entangled states. The rank of the given tensor representation for $\ket{\psi}$ is denoted by $(\chi_1, \chi_2, \ldots , \chi_N)$. Throughout the manuscript we choose $\chi_n = \chi$ for all $n$, reducing the number of hyperparameters. The choice of $\chi$, hereafter referred to as the tensor ring bond, for a specific problem significantly determines the representation ability and therefore performance of the algorithm. Each tensor in the the proposed TR representation is a third order tensor with a dimension of $\chi \times \chi \times 2$. The exponential reduction in storage complexity can be observed where a quantum state is represented by $2^N$ parameters, its TR approximation can be represented using only $2N\chi^2$ parameters. The approximation for a typical initialization for VQAs i.e., $\ket{0}^{\otimes{N}}$ can be easily computed to be a tensor ring with each tensor of dimension $\chi \times \chi \times 2$ where the value of the tensor is 1 at the index (1,1,1) and 0 elsewhere, represented by $\mathbbm{1}_{(1,1,1)}$. However, if a different initialization is to be chosen, constructing an approximation may not be as straightforward but efficient algorithms for TR decomposition have been studied at length in \cite{zhao2016tensor}.

\begin{figure}[!t]
 \centering
 \includegraphics[keepaspectratio, width = 0.5\textwidth]{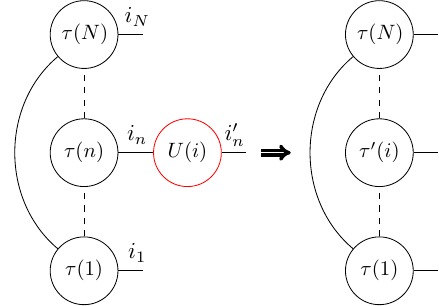}
 \caption{Tensor Ring transformation with a single qubit gate}
 \label{fig:tr_single}
\end{figure}

While a TR can represent a quantum state, it would also need to be transformed by parameterized rotations in order to function as specified in VQAs. Given the assumption of utilizing only single qubit gates and $CX$ gates in order to simplify the parameter shift rule, it would be sufficient to study the transformations of the TR corresponding to the aforementioned gate set. Unitary transformations of single qubits are represented by a $(2 \times 2)$ matrix which is a 2nd order tensor. The matrix multiplication associated can be implemented by contracting the unitary tensor along the free edge of the tensor corresponding to a qubit as specified in the following equation:

\begin{equation}\label{eq:nine}
    \tau'(n)_{r_{n-1} r_n}^{i'_n} = \sum_{i_n}U_{i'_n i_n} \tau(n)_{r_{n-1} r_n}^{i_n}
\end{equation}

$U_{i'_n i_n}$ is the 2nd order tensor with indices $i'_n$ and $i_n$ corresponding to the unitary matrix acting on $n$-th qubit which is contracted along the edge $i_n$ with the $n$-th tensor denoted by $\tau(n)$ spanning the indices $r_{n-1}, r_{n}$ and $i_n$, resulting in the new tensor $\tau'(n)_{r_{n-1} r_n}$. Note that the transformation associated with a single qubit rotation (visually illustrated in Fig \ref{fig:tr_single}) does not alter the structure of the tensor ring preserving the storage complexity.

Two qubit rotations like $CX$ however, can change the tensor ring structure increasing the storage complexity. In order to alleviate this, we use truncated singular value decomposition with the enlarged tensor to break it down to two tensors of the original smaller size. Say a two qubit gate $U \in \mathbb{R}^{4\times4}$ is to be applied to the adjacent qubits $m$ and $n$ (including the circular entanglement). We begin by contracting the two tensors $\tau(m)_{r_{m-1} r_m}^{i_m}$ and $\tau(n)_{r_{n-1} r_{n}}^{i_n}$ along their shared index $r_m = r_{n-1}$ to compute a new tensor:

\begin{equation}\label{eq:ten}
    M_{r_{m-1} r_{n}}^{i_m i_{n}} = \sum_{r_m} \tau(m)_{r_{m-1} r_m}^{i_m} \tau(n)_{r_{n-1} r_{n}}^{i_{n}}
\end{equation}

The two qubit gate $U$ is then reshaped into the tensor $U_{i'_m i'_{n} i_m i_{n}}$ and multiplied with the tensor $M_{r_{m-1} r_{n}}^{i_m i_{n}}$ along the shared edges:

\begin{figure*}[!t]
 \centering
 \includegraphics[keepaspectratio,width = 0.8\textwidth]{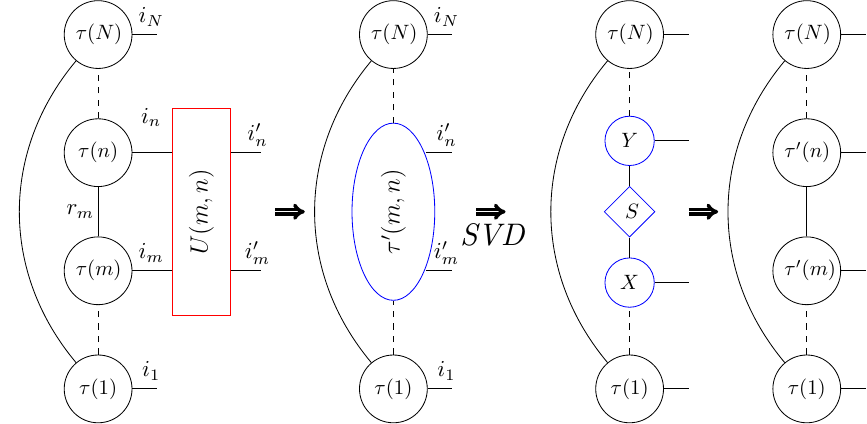}
 \caption{Tensor ring transformation with a two qubit gate}
 \label{fig:tr_two}
\end{figure*}

\begin{equation}\label{eq:eleven}
(\tau')_{r_{m-1} r_{n}}^{i'_m i'_{n}} = \sum_{i_m i_{n}} U_{i'_m i'_{n} i_m i_{n}}  M_{r_{m-1} r_{n}}^{i_m i_{n}}
\end{equation}

The resultant tensor is reshaped into a matrix of shape $(i'_m \times r_{m-1}) \times (i'_{n} \times r_{n})$ whose singular value decomposition is performed as follows:

\begin{equation}\label{eq:twelve}
(\tau')_{i'_m \times r_{m-1}}^{i'_{n} \times r_{n}} = \sum_{r_m} X_{r_{m-1} r_m}^{i'_m} S_{r_m} Y_{r_{n-1} r_{n}}^{i'_{n}}
\end{equation}

where the orthogonal vectors of $\tau'$ populate the matrices $X$ and $Y$ whereas $S_{r_m}$ is a diagonal matrix with the singular values. Since we assume a constant TR bond $r_m = \chi$ and we know the dimensionality of $i$ to be 2 ( the free indices span the quantum state), in this case, $\tau'$ has $2\chi$ singular values. $S_{r_m}$ is truncated resulting in a new diagonal matrix $S'_{r_m}$ with only the largest $\chi$ values remaining. We also truncate $X$ and $Y$ accordingly to keep only the orthogonal vectors corresponding to the remaining singular values. We compute products of the matrices $X,Y$ and $S$ as follows to make up the new tensors at the sites $m$ and $n$ of the tensor ring. Note that while this method can only work with two qubit gates acting on adjacent qubits,this can be extended to a generic circuit using SWAP gates.

\begin{equation}\label{eq:thirteen}
\tau'(m)_{r_{m-1} r_m}^{i'_m} = X_{r_{m-1} r_m}^{i'_m} S'_{r_m}
\end{equation}

\begin{equation}\label{eq:fourteen}
\tau'(n)_{r_{n-1} r_{n}}^{i'_{n}} =  Y_{r_{n-1} r_{n}}^{i'_{n}}
\end{equation}

Following the procedure specified, the resulting tensor ring would culminate with the same structure and dimensionality as before the procedure, preserving the storage complexity after each application of a two qubit rotation. It is to be noted, the specified operations at worst scale at $O(\chi^3)$, and without this approximation, the dimensionality of the tensor network approximation scales exponentially in the number of two-qubit rotations or the depth of the circuit, therefore increasing the computational complexity. Different stages of the two qubit rotation procedure with a TR is demonstrated in Figure \ref{fig:tr_two}.

Given that an ansatz has been chosen for a variational algorithm (assuming the conditions of only constructing a circuit with parameterized single qubit gates and $CX$ gates), it can be represented as a set of gates denoted by $\mathbbm{U}$, ordered by their position in the circuit i.e. a gate that is applied first to the quantum gate is placed at the beginning of the set, with the single qubit gates parameterized by $\theta_t$. The final quantum state produced by the circuit can be approximated by a tensor ring that is initialized as $\mathbbm{1}_{(1,1,1)}$ and transformed with each gate in $\mathbbm{U}$ as specified in the procedure in the preceding paragraphs. In order to compute the expected energy  with respect to the final quantum state, it must be decomposed into its linear sum of the expected energy of the unitary components of the Hamiltonian composed of Pauli matrices.

\begin{equation}\label{eq:fifteen}
    \bra{\psi(\theta)}\mathcal{H}\ket{\psi(\theta)} = \sum_{i,j} w_{i,j} \bra{\psi(\theta)}Z_iZ_j\ket{\psi(\theta)}
\end{equation}

\begin{figure}[!t]
 \centering
 \includegraphics[keepaspectratio,width = 0.4\textwidth]{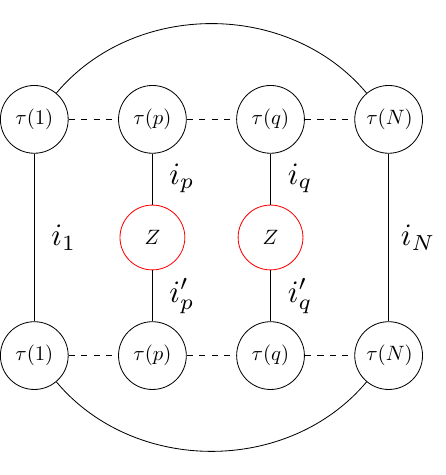}
 \caption{Evaluating the expected energy with respect to a quantum state using the TR approximation}
 \label{fig:tr_exp}
\end{figure}

We propose to compute the expected energy with respect to a component $Z_pZ_q$ using the TR representation by the application of single qubit Pauli Z gate at sites $p$ and $q$ and contracting it with the ring before the Z transformations along the edges that span the quantum Hilbert space (See Fig \ref{fig:tr_exp}).

\begin{equation}\label{eq:sixteen}
    \tau'(\theta)_{i_1\ldots,i'_p,\ldots,i'_q,\ldots i_N} = \sum_{i_p, i_q} Z_{i_p}^{i'_p} Z_{i_q}^{i'_q} \tau(\theta)_{i_1\ldots,i_p,\ldots,i_q,\ldots i_N}
\end{equation}

\begin{equation}\label{eq:seventeen}
    \bra{\psi(\theta)}Z_p Z_q\ket{\psi(\theta_t)} = \sum_{i_1,i_2,\ldots i_N} \tau'(\theta)_{i_1, i_2 \ldots i_N} \tau(\theta)_{i_1, i_2 \ldots i_N}
\end{equation}

In the equations above, $\tau(\theta)$ represents the final state produced by the ansatz $\mathbbm{U}$ parameterized by $\theta$ approximated by a TR and $\tau'(\theta)$ is produced after the Pauli Z transformations on the final state. Note that the indices $i'_p$ and $i'_q$ in $\tau'(\theta)$ have been renamed to $i_p$ and $i_q$  for a simplified representation. When computing the expected value, the order of the contractions becomes crucial to the computational complexity but it has been established \cite{orus2014practical} that it can be computed effectively in $O(N\chi^3)$ steps. The total procedure to compute the expected value has been presented in a more compact form in Algorithm \ref{algo:two}. We utilize this algorithm to evaluate the gradients of the variational quantum eigensolver by computing the expected energy of the two circuits with shifted parameters as shown in Algorithm \ref{algo:three}. The gradients are then used to update the weights of the variational parameters in the same manner as the naive VQE.

\subsection{Complexity}\label{sec:3-2}

\begin{figure*}[!t]
\centering
	\subfloat{\includegraphics[keepaspectratio,width=0.5\textwidth]{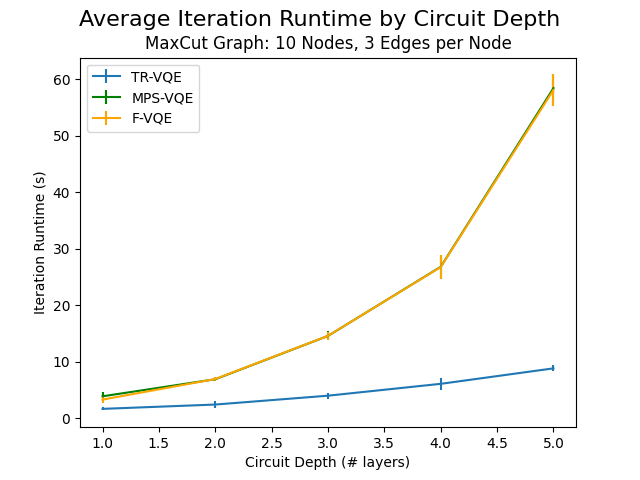}\label{fig:depth-rt}}
	\hfill
	\subfloat{\includegraphics[keepaspectratio,width=0.5\textwidth]{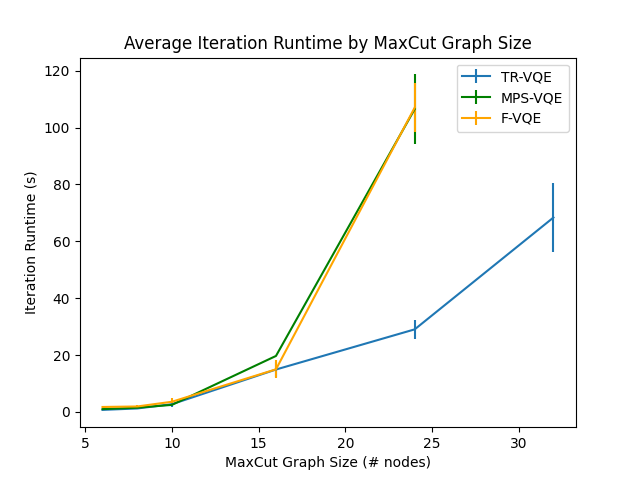}\label{fig:qubit-rt}}

\caption{Runtime for each iteration of the optimization with (Left) varying circuit depth or number of layers  and (Right) varying graph size or number of qubits across F-VQE, MPS-VQE and TR-VQE }%
\label{fig:rt}%
\end{figure*}

In terms of memory, we note that we construct and manipulate only a tensor ring with $N$ tensors corresponding to $N$ qubits which grows at the scale of $O(N\chi^2)$ as opposed to the $O(2^N)$ for the full quantum state. Zhou et al. \cite{zhou2020limits} establishes that the tensor network bond $\chi$ can be chosen to be sufficiently low in order to simulate a noisy quantum computer at a linear computational complexity in the number of qubits $N$ and circuit depth $D$ (defined as the number of repeating parametrized blocks). Parameter shift rule, popularized for its ability to compute the gradients on a quantum computer, evaluates the gradients by computing the expectations with shifted weights.However, computing the expected values with an additive error $\epsilon$ requires a many-fold implementation of the same circuit generally in the order of $O(1/\epsilon^2)$ which adds to the statistical noise. The proposed method can compute each gradient classically with a single iteration of two circuits each of which scales as $O(ND\chi^3)$ with an error rate controlled by $\chi$. The error rate introduced by the truncation decreases with an increasing bond dimension $\chi$ and generally saturates at a finite value in the order of $10^{-2}$ per two qubit gate for circuits with large $N$ and $D$. This is in contrast to the error rate on a quantum computer characterized by the fidelity per two qubit gate which exponentially decays in the overall number of gates in the circuit \cite{zhou2020limits}. The finite fidelity per gate allows us to scale the proposed algorithm in circuit depth and qubits for larger applications. Automatic differentiation (AD), a tool prevalent in classical machine learning literature and applications, grows at least as fast as the forward pass of the network in terms of computational complexity. This indicates that classically computing the gradients of VQE by AD scales exponentially as it would for classically computing the energy expectation of a circuit. It must be noted that the proposed method of tensor ring transformations can be used with AD as well, which again provides an exponential speedup in $N$ and $D$.

%% file: s4_experiments.tex
\section{Experiments} \label{sec:4}

To demonstrate the runtime performance and accuracy of the TR-VQE presented in Algorithm \ref{algo:three}, we compare several instances of training TR-VQE for MaxCut problem with Filtering VQE (F-VQE) \cite{amaro2022filtering} and naive VQE implemented on the Qiskit framework (MPS-VQE). Both the benchmarks use a non-noisy MPS representation to simulate the quantum computations from the circuit as formulated in \cite{vidal2003efficient, markov2008simulating} and the F-VQE is additionally implemented with an identity filter to equate the number of parameters in all the experiments. A sampling noise is introduced in the implementation of MPS-VQE and F-VQE to compute the expected values from the circuit. As discussed before, MPS-VQE is expected to compute more accurate gradients than TR-VQE owing to the induced noise in the proposed TR representation. Therefore MPS-VQE converges faster, however takes longer runtimes per iteration because the tensor sizes in MPS-VQE increase with circuit depth. F-VQE additionally implements filtering operators to change the optimization landscape thereby improving the training convergence. Amaro et al. \cite{amaro2022filtering} claims that the inclusion of filtering operators leads to a faster and more reliable convergence to the optimal solution. This improvement, however, is dwarfed with larger circuits with more number of qubits (Readers can refer to \cite{amaro2022filtering} for additional details on the implementation of F-VQE). We further collected data on TR-VQE to analyze how internal configurations, namely bond rank, and graph size, i.e., number of qubits affect the performance relative to filtering and naive VQE. All of the graphs used were randomly generated with two to three edges per node, and uniformly distributed weights (between 1 to 10) and edge pairs. We use the same circuit ansatz for all experiments, with an initial parameterized layer of $R_y$ gates on all qubits and a variational block repeated $D$ times, where $D$ represents the circuit depth. Each variational block contains a set of circular $CX$ or $CNOT$ gates followed by parameterized $R_y$ gates on all qubits followed by another set of $CX$ and $R_y$ gates. The circuit depth and the tensor ring rank is set to 1 and 10 respectively for all experiments, unless otherwise specified.
\begin{figure*}[!t]
\centering
	\subfloat{\includegraphics[keepaspectratio,width=0.5\textwidth]{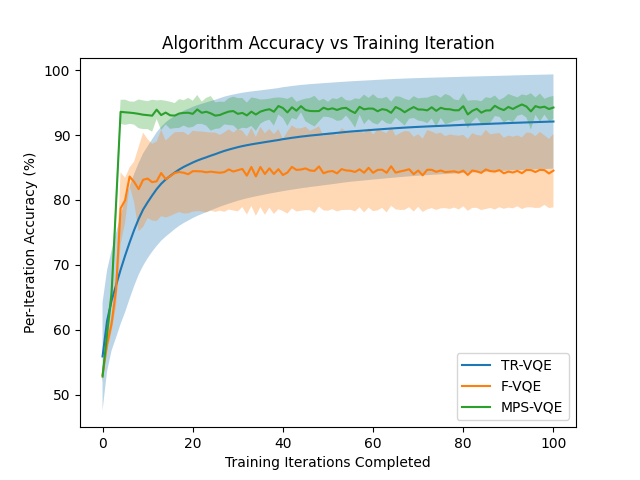}\label{fig:acc}}
	\hfill
	\subfloat{\includegraphics[keepaspectratio,width=0.5\textwidth]{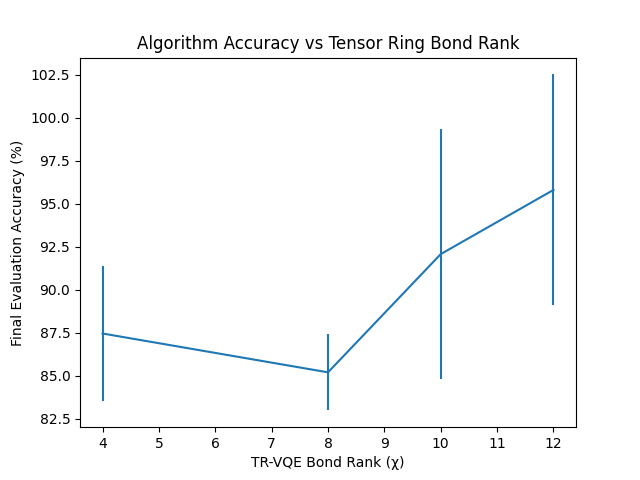}\label{fig:trrank}}

\caption{(Left) Plot depicts the improvement of approximation ratio with the iterations for TR-VQE, F-VQE and MPS-VQE. A statistical noise is introduced in MPS-VQE and F-VQE with the expectations sampled over 1000 shots (Right) Illustration of varying accuracy with different tensor ring bond rank}%
\label{fig:acc_plots}%
\end{figure*}

\begin{figure*}[!t]
\centering
	\subfloat{\includegraphics[keepaspectratio,width=0.35\textwidth]{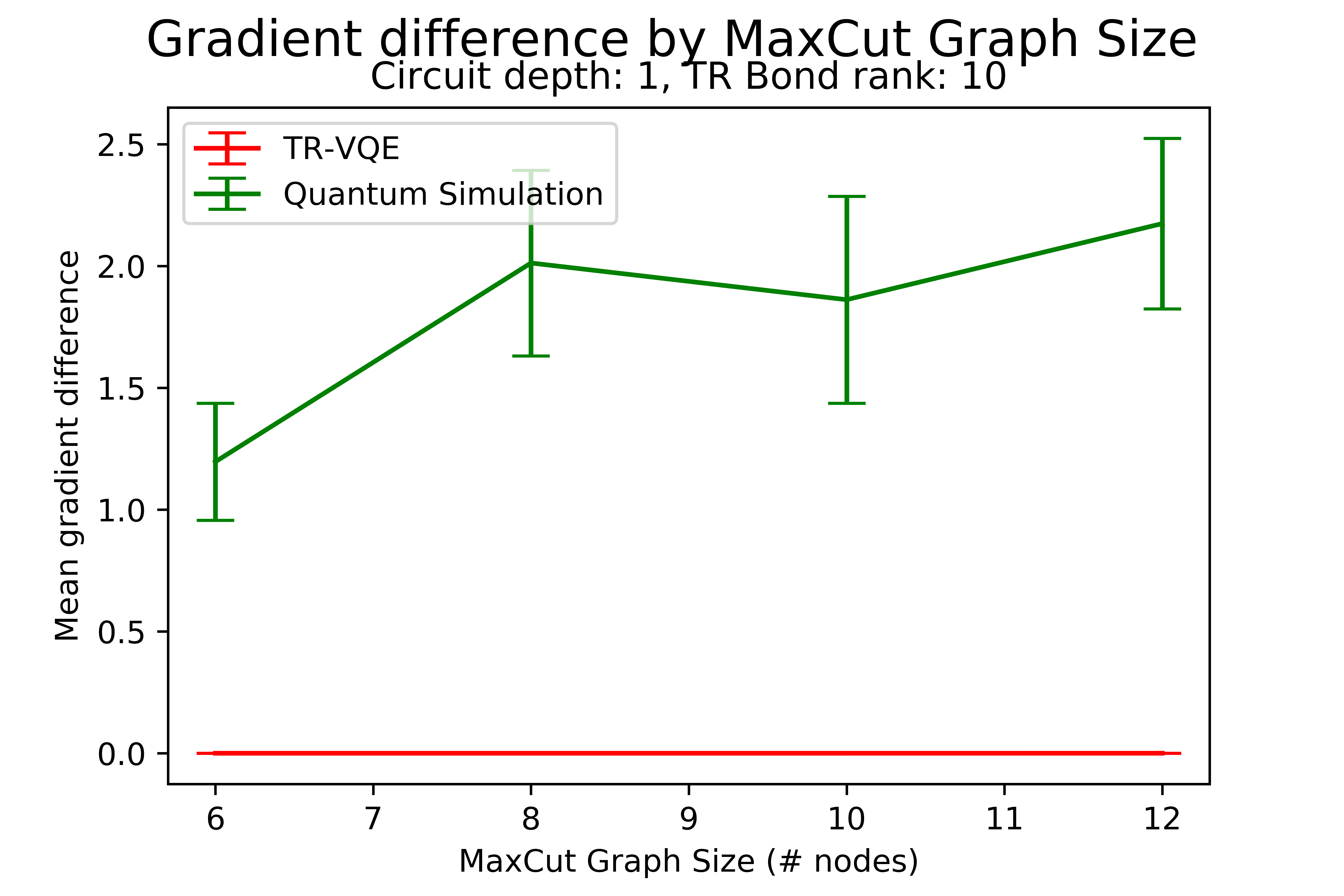}\label{fig:grad1}}
	\subfloat{\includegraphics[keepaspectratio,width=0.35\textwidth]{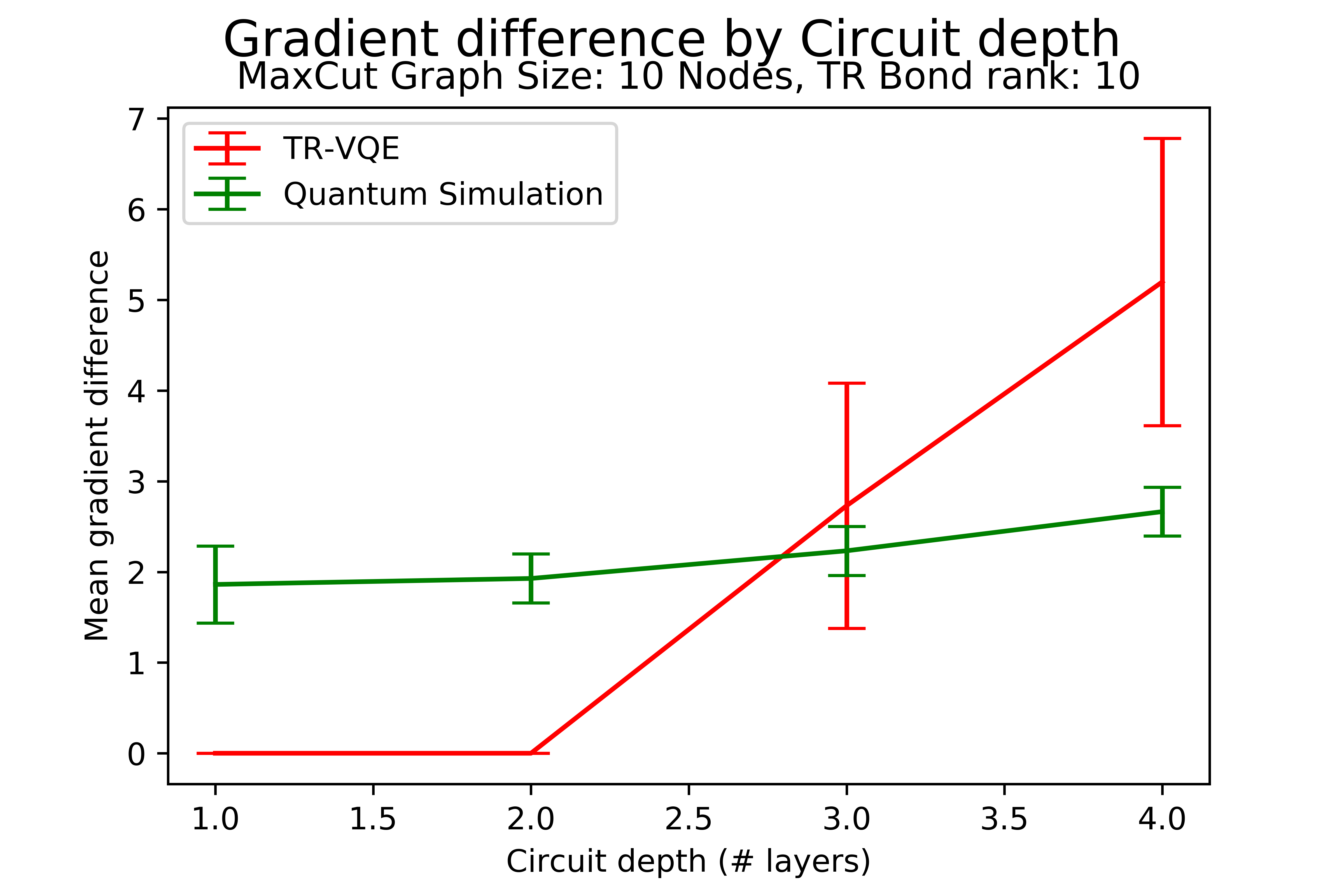}\label{fig:grad2}}
    \subfloat{\includegraphics[keepaspectratio,width=0.35\textwidth]{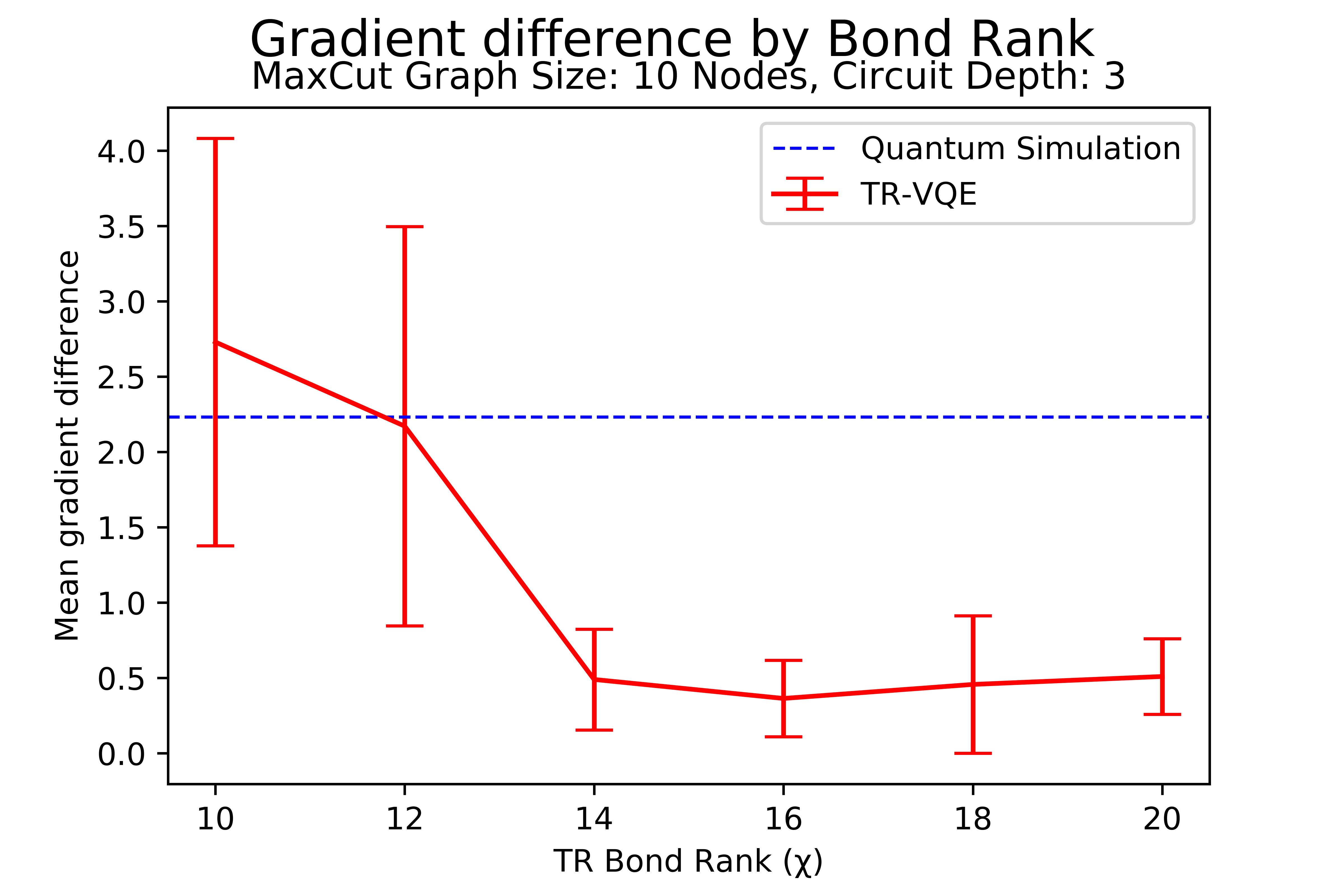}\label{fig:grad3}}

\caption{(Left) Plot depicts the mean gradient distance of TR-VQE and noisy quantum simulation 
 over problems of multiple graph sizes. The circuit depth and TR bond rank are set to 1 and 10, respectively. (Middle) Mean gradient distances of TR-VQE and noisy quantum simulation are plotted against the circuit depth. Graph size and bond rank are both set to 10. (Right) Mean gradient distance with increasing bond rank. The dotted line represents the same value from noisy quantum simulation. The circuit depth and graph size are set to 3 and 10, respectively.}%
\label{fig:grad_plots}%
\end{figure*}



\begin{table*}[!t]
\centering
\resizebox{0.5\textwidth}{!}{
\begin{tabular}{|c|ccc|}
\hline
\textbf{Graph Size} & \textbf{TR-VQE} & \textbf{MPS-VQE} & \textbf{F-VQE} \\ \hline
6                   & 97.68\%         & 83.26\%          & 94.11\%        \\
8                   & 93.47\%         & 97.14\%          & 91.55\%        \\
10                  & 93.17\%         & 94.26\%          & 84.44\%        \\
16                  & 90.70\%         & 94.33\%          & 93.83\%        \\ \hline
\end{tabular}}
\caption{Optimal approximation ratio 
 (Equation \ref{eq:approx} averaged across 10 different graphs for each graph size, trained over 100 iterations) of TR-VQE compared against those of MPS-VQE and F-VQE for various graph sizes. An additional sampling noise in the order of $10^{-1.5}$ has been considered for MPS-VQE and F-VQE }\label{tab:results}
\end{table*}

Figure~\ref{fig:rt} indicates how each of the three algorithms performs in terms of iteration runtime across randomly generated graphs of varying sizes and different circuit ansatz. The results for each algorithm were averaged across 10 initializations each with multiple unique MaxCut graphs of fixed size. For MPS-VQE and F-VQE, the number of shots used in the Hamiltonian evaluation was increased quadratically in graph size. Across varying graph sizes, TR-VQE’s per-iteration runtime, computed as the time taken for computing the expected value of the Hamiltonian and updating the parameters from the evaluated gradients, is faster than both filtering and non-filtering VQE with smaller graphs and by extension, smaller number of qubits. As illustrated in Figure \ref{fig:qubit-rt}, the iteration runtimes of TR-VQE consistently improve by a large margin over the benchmarks when the number of qubits are increased. Figure \ref{fig:depth-rt} demonstrates the iteration runtime of each algorithm with increasing circuit depths for a graph with 10 nodes. TR-VQE again shows a significant improvement in runtime compared to MPS-VQE and F-VQE with increasing number of layers. The results from both the experiments are compatible with the theoretical claims of improved runtime complexity as discussed in Section \ref{sec:3-2}. The runtime speedup can be attributed to the consistent rank and tensor sizes irrespective of the circuit depth whereas in the naive MPS based approach, the tensor sizes increase with the circuit depth.


On the other hand, TRVQE performs with near-equivalent accuracy to the other algorithms, despite the runtime speedup. Figure~\ref{fig:acc} displays per-iteration accuracy for the algorithms, averaging data from 10 runs on various randomly generated graphs with a fixed size of 10 nodes. The accuracy was compared using the approximation ratio at each iteration computed as defined in equation \ref{eq:approx}.

The resulting data from Figure \ref{fig:acc} indicate that TR-VQE performs similar to F-VQE in terms of accuracy, diverging on average by no more than 3\% at any point during training. When extended to variable graph sizes, TR-VQE once again performs on par or better than the alternative algorithms. The data in Table~\ref{tab:results} was collected using a TR-VQE bond rank of 10 and 1000 shots per circuit evaluation for MPS-VQE and F-VQE. Excluding an outlier at small graph sizes due to instability, MPS-VQE performed the most accurately due to the availability of more information, albeit at the cost of larger runtime. However, TR-VQE followed closely behind, with a large but inconsistent gap in accuracy between it and the least accurate F-VQE algorithm.


We also plot the approximation ratio of TR-VQE with varying TR bond rank and it is to be noted that TR-VQE performs almost as good as MPS-VQE at ranks as low as 12, indicating that an exponential speedup can be achieved at smaller ranks, improving the storage complexity. All experiments including the benchmarks see a wide variance in terms of accuracy with larger graph sizes due to a phenomenon called the barren plateau effect \cite{mcclean2018barren} which is informally defined as the impaired performance due to the exponential flattening of loss landscape in the number of qubits. Martin et al. \cite{martin2023barren} demonstrate that barren plateau effect persists in quantum MPS circuits and therefore we can surmise that Tensor ring circuits, as an extension of MPS, will face a similar challenge in training.

To assess the accuracy of approximate gradients, we employ the $l^2$-norm to compare gradients obtained from state vector simulations and those generated using the TR-VQE method. The mean gradient distance, computed as the average norm difference across 500 randomly selected points on the optimization landscape, is used as a metric. We compare this metric with values obtained from noisy simulations that emulate the gradients on an actual quantum computer using noise models from the $ibm\  montreal$ machine. We examine the mean gradient distance for various circuit depths and graph sizes.

Figure \ref{fig:grad_plots}(Left) illustrates that the gradients produced by the TR-VQE method closely resemble those obtained from exact state vector simulations, with almost negligible differences. In contrast, gradients derived from quantum simulation deviate significantly from the exact gradients, a trend that becomes more pronounced as the number of qubits increases, as expected. As shown in Figure \ref{fig:grad_plots}(Middle), TR-VQE's effectiveness diminishes with higher circuit depths due to the cumulative impact of two-qubit gates. However, this performance decline can be mitigated by increasing the tensor rank, as demonstrated in Figure \ref{fig:grad_plots}(Right). In conclusion, gradients computed from approximate classical simulations can achieve accuracy comparable to those obtained from quantum computers. Consequently, they can be a valuable addition to the optimization process in hybrid algorithms.



%% file: s5_conclusion.tex
\section{Conclusion} \label{sec:5}
This work proposes a novel technique for combinatorial optimization problems with Variational Quantum Eigensolvers by approximating the circuit computations with noisy tensor ring contractions. The proposed algorithm uses parameter shift rule to evaluate the gradients used to update the variational parameters, but computes the expected values of the shifted circuits using tensor ring approximation. The computational complexity of circuit evaluation grows linearly in the number of qubits and the circuit depth which offers a quadratic speedup over the perfect classical simulation. Evaluating gradients using TR-VQE can also eliminate the additive error present in circuit computations on quantum computers. We validate the algorithm by implementations on several instances of Max-Cut problem and compare with algorithms that use the full state information. The results demonstrate the vast improvement in runtime with respect to the number of qubits and circuit depth validating the complexity analysis at a minor cost of accuracy.